\documentclass[12pt]{article}

\RequirePackage[OT1]{fontenc}
\RequirePackage{amsthm,amsmath,amssymb,bm,bbm,stackrel,enumerate}
\RequirePackage{natbib}
\RequirePackage[colorlinks,citecolor=blue,urlcolor=blue]{hyperref}

\usepackage{pdfpages}

\usepackage{mathtools}

\usepackage{url}

\usepackage{fullpage}
\usepackage{graphicx}

\usepackage{exscale}
\usepackage{relsize}
\usepackage{graphicx} 
\usepackage{rotating}

\usepackage{caption}
\usepackage{subcaption}
\usepackage{verbatim}

\usepackage{makecell}
\usepackage{cleveref, autonum}
\usepackage{booktabs}
\usepackage{multirow}
\usepackage{pdflscape}
\usepackage{ltablex}
\usepackage{authblk}

\title{\bf Function-on-Scalar Quantile Regression with Application to Mass Spectrometry Proteomics Data}
\author[*]{Yusha Liu}
\author[*]{Meng Li}
\author[$\dag$]{Jeffrey S. Morris}
\affil[*]{Department of Statistics, Rice University}
\affil[$\dag$]{Division of Biostatistics, University of Pennsylvania}

\begin{document}

\def\spacingset#1{\renewcommand{\baselinestretch}%
{#1}\small\normalsize} \spacingset{1}

\maketitle

\begin{abstract}
Mass spectrometry proteomics, characterized by spiky, spatially heterogeneous functional data, can be used to identify potential cancer biomarkers. Existing mass spectrometry analyses utilize mean regression to detect spectral regions that are differentially expressed across groups. However, given the inter-patient heterogeneity that is a key hallmark of cancer, many biomarkers are only present at aberrant levels for a subset of, not all, cancer samples.  Differences in these biomarkers can easily be missed by mean regression, but might be more easily detected by quantile-based approaches. Thus, we propose a unified Bayesian framework to perform quantile regression on functional responses. Our approach utilizes an asymmetric Laplace working likelihood, represents the functional coefficients with basis representations which enable borrowing of strength from nearby locations, and places a global-local shrinkage prior on the basis coefficients to achieve adaptive regularization. Different types of basis transform and continuous shrinkage priors can be used in our framework. A scalable Gibbs sampler is developed to generate posterior samples that can be used to perform Bayesian estimation and inference while accounting for multiple testing. Our framework performs quantile regression and coefficient regularization in a unified manner, allowing them to inform each other and leading to improvement in performance over competing methods as demonstrated by simulation studies. We also introduce an adjustment procedure to the model to improve its frequentist properties of posterior inference. We apply our model to identify proteomic biomarkers of pancreatic cancer that are differentially expressed for a subset of cancer patients compared to the normal controls, which were missed by previous mean-regression based approaches. Supplementary materials for this article are available online.
\end{abstract}

\noindent%
{\it Keywords:}  Functional data analysis, Functional response regression, Quantile regression, Bayesian hierarchical model, Global-local shrinkage, Proteomic biomarker
\vfill

\newpage

\spacingset{1.45}

\section{Introduction} \label{introduction}
\subsection{Mass Spectrometry Proteomics} \label{spectra}
The rapid advancement of molecular biotechnology has led to the ability to make large-scale molecular measurements using high-throughput technologies at various molecular resolution levels, including DNA, mRNA, epigenetic, metabolite, and protein levels. DNA and mRNA have been most frequently studied, largely because nucleotide sequences are easier to study and analyze in nature than proteins and metabolites. However, it is proteins, rather than DNA or messenger RNA, that play a fundamental functional role in the molecular processes underlying various diseases including cancer. As a result, there is great interest in studying proteins directly and identifying proteomic biomarkers of cancer that can potentially be used for early detection, new drug target identification and precision medicine strategies.

Mass spectrometry is an analytical technique to survey a large number of different proteins, peptides, or metabolites in a biological sample by first ionizing the particles from the sample, then separating the ions based on their mass-to-charge ratio, and detecting the ions and assembling them into a mass spectrum for each sample. Commonly used ionization techniques for solid and liquid biological samples include MALDI (matrix assisted laser desorption and ionization) and ESI (electrospray ionization), and popular mass analyzers which separate charged particles include TOF (time-of-flight) analyzer and quadrupole mass analyzers. Regardless of the ionization and separation techniques used, the resulting mass spectrum is a highly spiky and irregular function with many peaks, with the spectral intensity $y(t)$ approximating the relative abundance of a protein or peptide with the mass-to-charge ratio of $t$ in the given biological sample. To further enhance its capability for protein identification and quantification, mass spectrometry is often used in tandem with liquid chromatography, which first separates the proteomic sample through an LC column over a series of elution times based on hydrophobicity or other physical properties before the mass spectrometry procedure, resulting in 2D mass spectrometry data (LC-MS) with one dimension representing elution time and the other dimension representing the mass-to-charge ratio~\citep{zhang2009review,liao2014new}.

\subsection{Interpatient Heterogeneity and Pancreatic Cancer Proteomic Markers} \label{motivation}

\begin{figure}[h]
    \centering
    \includegraphics[scale=0.32]{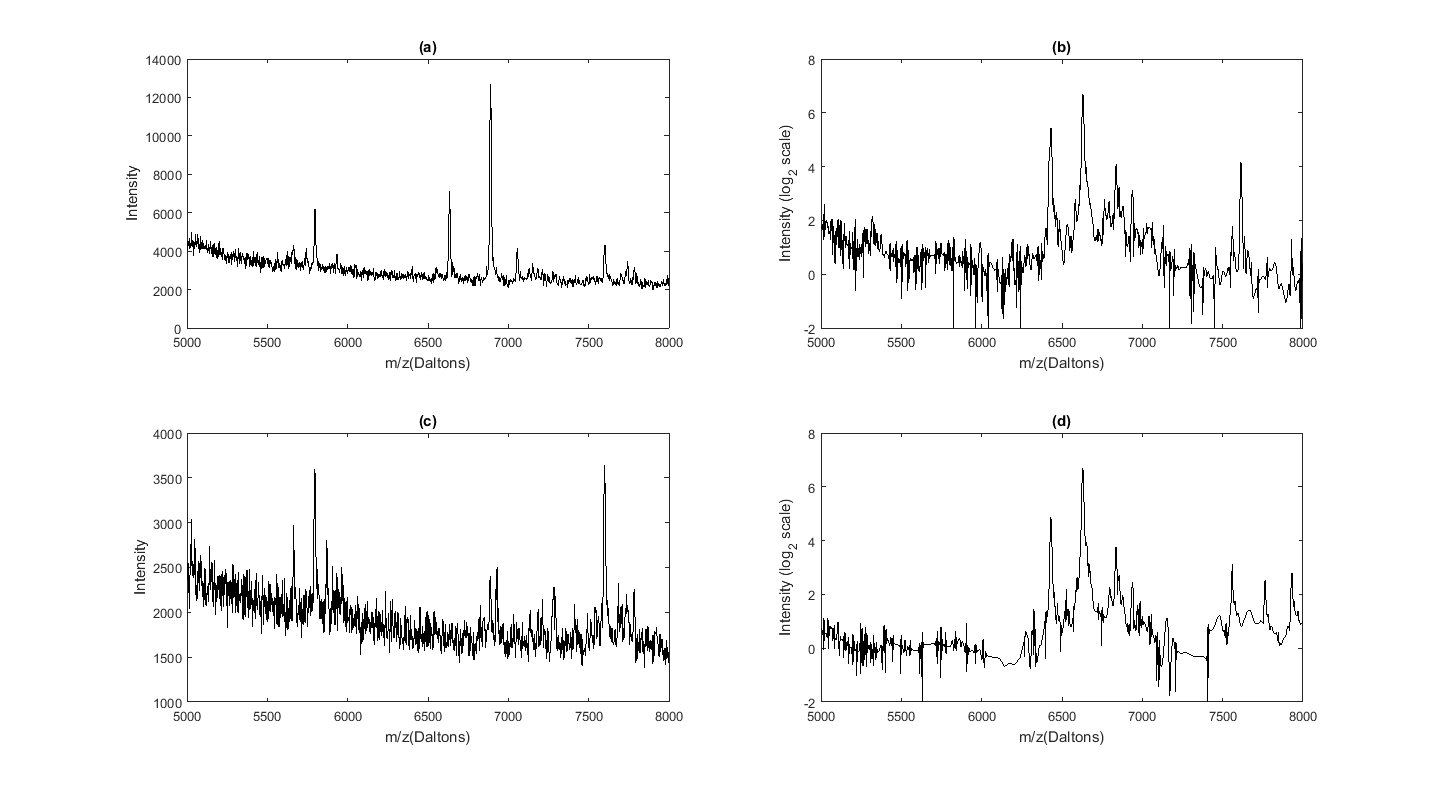}
    \caption{\textbf{\textit{Sample spectra from the pancreatic cancer dataset.}} The first column shows the raw spectrum of a normal control (a) and a cancer patient (c) randomly chosen from the pancreatic cancer MALDI-TOF mass spectrometry dataset. The second column displays the corresponding spectra of the normal control (b) and the cancer patient (d) after preprocessing, which includes baseline correction, normalization, denoising and $\log_2$ transformation.}  
    \label{fig:figure1}
\end{figure}

At the University of Texas M.D. Anderson Cancer Center, a study was conducted using MALDI-TOF to discover potential proteomic markers of pancreatic cancer. In this study, researchers collected the blood serum samples from $139$ pancreatic cancer patients and $117$ normal controls and ran them on a MALDI-TOF mass spectrometer to produce a mass spectrum for each sample~\citep{koomen2005plasma,morris2008bayesian}. The left column of Figure~\ref{fig:figure1} displays the raw spectrum of a pancreatic cancer patient and a normal control from this dataset, which demonstrates the highly spiky and irregular nature of mass spectrometry data. We also provide plots of a random sample of individual spectra in Figure S1 in the supplementary to give readers an idea of the characteristics of these functional data. The primary goal of this study is to identify proteins, represented by spectral regions, with differential abundance between pancreatic cancer and normal samples, and potentially useful as diagnostic, prognostic, or predictive biomarkers.

Classic approaches to analyzing mass spectrometry data depend on first performing peak detection, and then only analyzing the detected locations and sometimes intensities of those peaks. For example, after applying a feature detection method to identify $m$ peaks for each of $N$ spectra, these can be put together into an $N\times m$ matrix and analyzed to find which of the $m$ features are associated with factors of interest (cancer/normal). While this two-step approach seems intuitive and reasonable, important proteomic differences across factors of interest might be missed if the feature detection procedure in the first step fails to detect peaks corresponding to the corresponding protein. An alternative to this feature extraction-based approach is to model the entire mass spectra as functional data using functional data analysis techniques. \citet{morris2012statistical} applied the wavelet-based functional mixed model introduced by~\citet{morris2006wavelet} to this pancreatic cancer dataset to identify differentially expressed regions between cancer and control in the range from $t=4,000$ to $t=20,000$ Daltons, and flagged approximately $50\%$ more significant spectral regions than the more commonly used peak detection approach, suggesting that the functional modeling approach can yield greater power for biomarker discovery. 

As is the case for nearly all mass spectrometry analyses, both of these feature extraction and functional data approaches utilize mean regression, in which the mean expression levels are compared across pre-defined groups.\\
However, given the interpatient heterogeneity that is a hallmark of cancer, many potentially useful proteomic biomarkers may have aberrant expression in only a small subset of cancer patients compared to the normals. The cancer-normal differences in these cases may not be apparent in the means, but would manifest in certain quantiles in the tail of the distribution.

\begin{sidewaysfigure}[ht]
    \centering
    \includegraphics[width=0.85\textwidth]{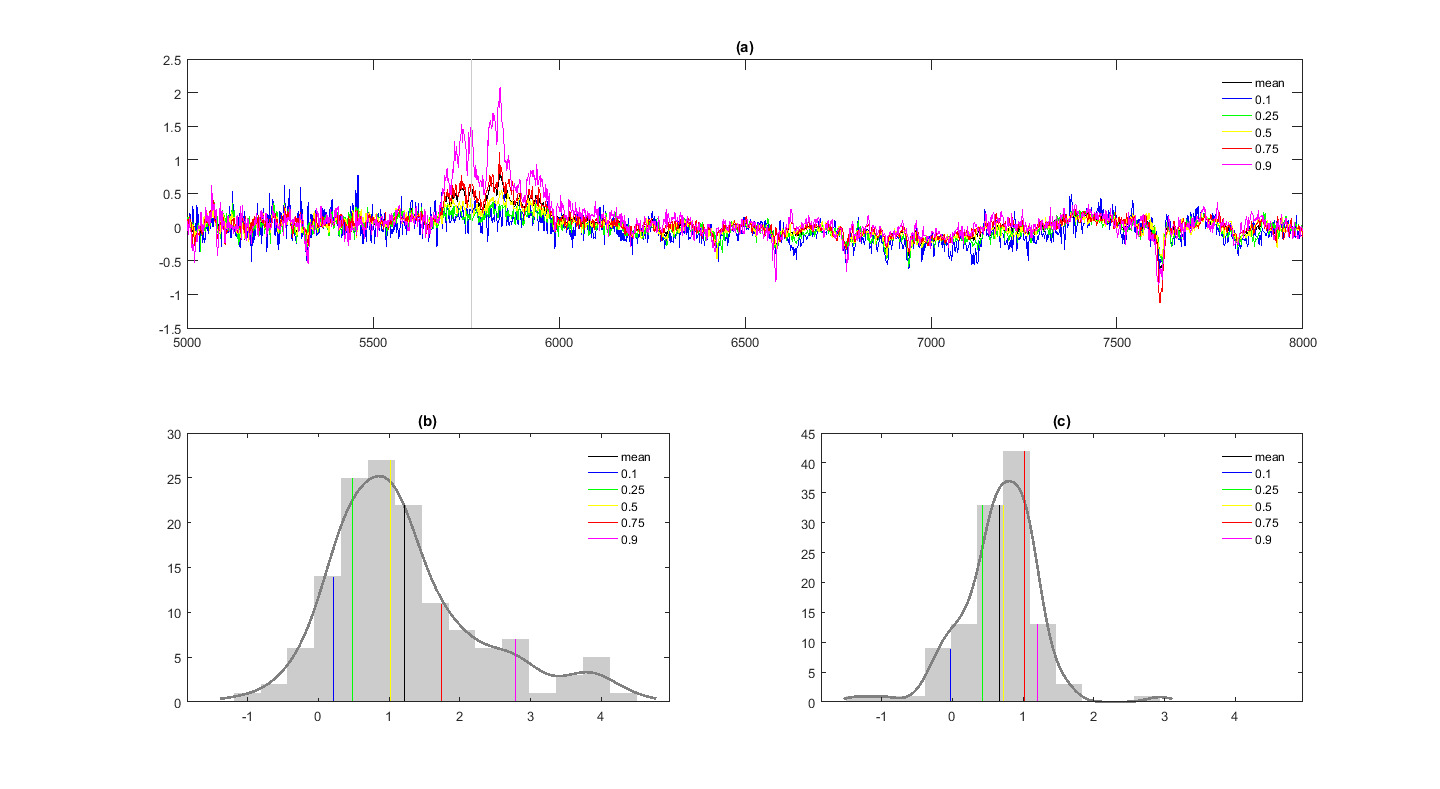}
    \caption{\textbf{\textit{Empirical comparisons of spectral intensities on the $\log_2$ scale between the cancer and normal cohorts at multiple quantile levels and mean.}} (a) shows the differences of spectral intensities of the sample mean and multiple quantile levels (denoted by different colors) in the spectral region [$5000$D, $8000$D] between the cancer and normal cohorts from the preprocessed pancreatic cancer dataset. The vertical line represents the spectral location $5764.1$D, where we observe a huge difference between the two cohorts in the $90$th sample quantiles. The histograms of the spectral intensities at $5764.1$D are shown respectively for the cancer cohort (b) and the normal cohort (c), with the sample mean and multiple quantile levels marked by vertical lines with different colors.}
    \label{fig:figure2}
\end{sidewaysfigure}

To explore this possibility, we computed the difference in the mean and sample quantiles between the cancer and normal groups for each spectral position between $5000$D and $8000$D in Figure~\ref{fig:figure2} (a). Note that in the region ($5700$D, $6000$D), there appear to be huge differences in the $90$th percentile in the upper tail, while there is little evidence of a difference in the median or mean. More closely inspecting one location at $5764.1$D, Figure~\ref{fig:figure2} (b) and (c) show a strongly right skewed pattern of the spectral intensity distribution for the cancer cohort and a slightly left skewed distribution with a similar mode for the normal cohort. This observation suggests that a small subset of pancreatic cancer patients have much higher protein levels than other patients and healthy controls at $5764.1$D, but mean or median regression might not be able to detect this important pattern. While these plots are suggestive of some difference, formal statistical methods are needed to assess these potential differences, and these methods need to account for the multiple testing problem inherent to these high-dimensional data. Our goal in this paper is to develop such methods.

\subsection{Literature Review and Contributions} \label{review}

Quantile regression, first introduced by \citet{koenker1978regression}, has been widely used in many application areas to study the effect of predictor variables on a given quantile level of the response, and can reveal important information about how the entire distribution of response varies with predictors in ways that might not be captured by mean regression. Traditionally, quantile regression is formulated as an optimization problem in which the regression coefficients are estimated by minimizing the check loss function~\citep{koenker2005quantile}. 

Recently, Bayesian quantile regression has gained a lot of attention, partly because posterior samples can be used to perform Bayesian inference on any model parameter in a straightforward manner. A great variety of likelihoods have been proposed to perform Bayesian quantile regression; see~\citet{lum2012spatial} and~\citet{yang2016posterior} for a comprehensive overview. Among them, \citet{yang2012bayesian} used the empirical likelihood in a Bayesian framework, making it possible to model multiple quantiles simultaneously and achieve efficiency gains through borrowing strength across quantiles, and established the frequentist asymptotic validity of posterior inference based on the empirical likelihood. \citet{xi2016bayesian} extended this approach to perform Bayesian variable selection in quantile regression by putting a spike-and-slab prior on the regression coefficients. In this paper, we chose to use asymmetric Laplace (AL) error distribution~\citep{yu2001bayesian} that has been widely adopted in Bayesian quantile regression \citep{geraci2006quantile, yue2011bayesian, lum2012spatial}, based on the fact that the maximization of an AL likelihood is equivalent to the minimization of the check loss function.

\clearpage

\clearpage

In the present context, we would like to perform quantile regression for each spectral location, which is a generalization consisting of quantile regression of functional responses on scalar predictors that we henceforth refer to as \textit{functional quantile regression}(FQR). \citet{kim2007quantile} introduced a varying coefficient model for quantile regression, which can model the effect of a continuous predictor on the conditional quantile of a scalar response nonparametrically, and there has been some extensions along this line of work including \citet{cai2008nonparametric, wang2009quantile, feng2016estimation}. There has also been recent work on scalar-on-function quantile regression, where the conditional quantile of a scalar response is modeled as an inner product of a functional predictor and an unknown coefficient function~\citep{Cardot+:05,ferraty2005,chen+muller2012,Kato:12,li2016inference}. However, to the best of our knowledge, there is a paucity of methods to perform FQR, i.e., function-on-scalar quantile regression. One approach would be to simply fit independent quantile regressions for each $t$, which is unbiased but expected to be inefficient since it does not borrow strength from nearby $t$ as is typical in functional data modeling approaches. As emphasized in a review of functional regression techniques in~\citet{morris2015functional}, most functional regression methods borrow strength across $t$ by using basis functions and penalization to induce smoothness and regularization in the functional coefficients. The functional linear array model proposed by~\citet{brockhaus2015functional} is a general framework for functional regression that can be used to perform FQR if the check loss function is used. However, as we will show by simulations, this framework's utilization of spline basis functions and global L$2$ penalization may not work well for complex, irregular functions like the mass spectrometry data here, and the FDboost fitting approach~\citep{fdboost2017} has scalability problems in this setting. New methods for performing functional response quantile regression are needed for such data.

We make the following contributions in this paper.  Motivated by the mass spectrometry dataset, we present a novel unified Bayesian FQR framework that is designed for complex, high-dimensional functional data that are sampled on a dense grid. Our proposed framework adopts AL distributed residual error functions, which lead to quantile regression on functional responses, and adaptively regularizes the functional regression coefficients using a basis representation with shrinkage priors on the corresponding basis coefficients. This framework is highly general in that any basis functions and computationally tractable shrinkage priors can be chosen, depending on the characteristics of the functional data to be analyzed. It is also easy-to-implement, given that basis transforms and hierarchical shrinkage priors are well-developed and frequently used tools in Bayesian modeling nowadays. In addition, this framework not only yields estimates, but also posterior samples that can be used to perform Bayesian inference on the regression coefficients while accounting for multiple testing over $t$. We develop a scalable Gibbs sampler to fit this fully Bayesian hierarchical model in an automated fashion with no tuning required. Our approach is computationally scalable and can handle functional responses observed on grids of hundreds to thousands. We apply our model to identify proteomic biomarkers of pancreatic cancer that are differentially expressed for a subset of cancer patients compared to the normal controls, which were missed by previous mean-regression based approaches.

We introduce the Bayesian functional quantile regression framework in Section \ref{model}, describe the procedures for posterior computation of our proposed model in Section \ref{Gibbs}, discuss posterior inference in Section \ref{inference}, and propose an adjusted version of our model to improve the frequentist properties of posterior inference in Section \ref{correction}. We conduct simulation studies to assess the performance of our model and compare to other alternatives in Section \ref{simulations}, apply our model to the motivating pancreatic cancer mass spectrometry dataset and discuss the findings in Section \ref{application}, and conclude the paper with a discussion in Section \ref{discussion}.

\section{Methods} \label{methods}
\subsection{Bayesian functional quantile regression (FQR) model} \label{model}
Suppose a sample of $N$ curves $\mathbf{Y}(t)=(Y_1(t),\dots, Y_N(t))'$ are observed on the same compact set \( \mathcal{T} \), and $\mathbf{X}$ is the $N\times p$ design matrix. For the $\tau$th quantile, the model we use to perform Bayesian functional quantile regression is given by
\begin{align} \label{eq:1}
\mathbf{Y}(t)=\mathbf{X}\mathbf{B}^{\tau}(t) + \mathbf{E}^{\tau}(t), 
\end{align}
where $\mathbf{B}^{\tau}(t)=(B_1^{\tau}(t),\dots, B_p^{\tau}(t))'$ is a vector of regression coefficient functions measuring the effect of covariates $\mathbf{X}$ on the $\tau$th quantile of response function $Y$ at position $t$, and $\mathbf{E}^{\tau}(t)=(E_1^{\tau}(t),\dots, E_N^{\tau}(t))'$ is a vector of residual error functions that follow asymmetric Laplace distribution $\text{AL}(0, \tau, \sigma(t))$ at position $t$, independently across positions and samples.

The probability density function of $\text{AL}(0, \tau, \sigma(t))$ is given by
$$ f(\epsilon|\mu, \tau, \sigma)=\frac{\tau(1-\tau)}{\sigma} \text{exp}\left[-\frac{\rho_\tau(\epsilon-\mu)}{\sigma}\right], $$
where $\rho_\tau(u)= u(\tau - \mathbbm{1}_{(u\le0)})$ is the check loss function. The $\tau$th quantile of the asymmetric Laplace distribution $\text{AL}(0, \tau, \sigma(t))$ is zero, therefore, model \eqref{eq:1} implies $Q_{\tau}(\mathbf{Y}(t)|\mathbf{X})=\mathbf{X}\mathbf{B}^{\tau}(t)$ for $\forall  t \in \mathcal{T}$, with $Q_{\tau}(\mathbf{Y}(t)|\mathbf{X})$ denoting the $\tau$th quantile of $\mathbf{Y}(t)$ conditional on $\mathbf{X}$, and $B_a^{\tau}(t)$ representing the partial effect of the covariate $a$ on the $\tau$th quantile of $\mathbf{Y}(t)$.
An asymmetric Laplace random variable $\epsilon$ can be represented as a scale mixture of normal distributions~\citep{reed2009partially}, i.e., \\[-12pt] 
$$ \epsilon \stackrel{d}{=} \frac{1- 2\tau}{\tau(1-\tau)} \xi + \sqrt{\frac{2\sigma\xi}{\tau(1-\tau)}} Z,$$
where $Z$ is a standard normal random variable and $\xi$ is an independent exponential random variable with mean $\sigma$. This representation allows the development of an efficient partially collapsed Gibbs sampler for Bayesian quantile regression as detailed in Section~\ref{Gibbs}.

To simplify notation, henceforth we omit the quantile level $\tau$ in the hierarchical modeling assumptions we make for the functional quantile regression coefficients $\mathbf{B}^{\tau}(t)$, with the understanding that the coefficients correspond to a particular choice of quantile $\tau$.

\textbf{Basis Representation and Shrinkage Priors:}. As is typical for functional regression methods, we will induce regularization in the functional coefficients $B_a(t)$ using a basis representation and penalization induced by sparsity priors.  For a given chosen finite basis representation $\{\phi_k(t), k=1,\ldots,K\}$, we specify a basis representation for $B_a(t)$, \\[-18pt]
\begin{align} \label{eq:2}
B_a(t)=\sum_{k=1}^{K}B^*_{ak} \phi_k(t). 
\end{align} 
Common choices of the basis functions include splines, functional principal components, Fourier bases and wavelets. 

As is typical in functional regression contexts \citep{morris2015functional}, appropriate regularization of basis coefficients $B^*_{ak}$ produces smoother and more regular estimates of the corresponding functional coefficients $B_a(t)$ that borrow strength across $t$. We choose to penalize the basis coefficients using a global-local shrinkage prior, which consists of a global shrinkage parameter whose prior has substantial mass near zero to handle noise effectively, and a local shrinkage parameter whose prior has a heavy tail to avoid over-shrinkage of signals ~\citep{polson2010shrink}. Global-local shrinkage priors have been widely used in Bayesian modeling these days, and some of them, including the horseshoe and the Dirichlet-Laplace prior, have been shown to possess desirable theoretical properties in the high-dimensional regression setting \citep{carvalho2010horseshoe, van2014horseshoe, bhattacharya2015dirichlet}. For extra flexibility in regularization, we group the basis functions $k=1,\ldots,K$ into regularization subsets $j=1,\ldots, J$, each containing $H_j$ basis functions such that $K=\sum_{j=1}^J H_j$.  This allows different sets of basis functions to experience different levels of shrinkage, which can lead to more adaptive regularization of $B_a(t)$.  For example, for wavelet bases, $j$ can index the wavelet scale, allowing higher and lower frequency wavelets to experience different levels of shrinkage.  For functional principal components analysis, the $H_j$ eigenfunctions that share the same $\left \lfloor{\log_{10}(\eta_k)}\right \rfloor$, where $\eta_k$ denotes the corresponding eigenvalue, can be grouped into the same regularization subset $j$, allowing the possibility that dimensions explaining a higher proportion of the functional variability may also be more important for representing the functional predictor $B_a(t)$ as well, and be allowed to experience less shrinkage.  

Given the regularization groups, a general global-local prior on the basis coefficients $B^*_{ajh}$, where the subscripts $j$ and $h$ index the regularization subset and basis function respectively, can be expressed as
\begin{align} \label{eq:3}
B^*_{ajh} \sim N(0,\lambda^2_{ajh}\psi^2_{aj}), \quad  \lambda_{ajh} \sim g_1, \quad  \psi_{aj} \sim g_2(\Theta_{aj}).  
\end{align}
This prior is comprised of a scale mixture of Gaussians, with a global shrinkage parameter $\psi^2_{aj}$ and local shrinkage parameter $\lambda^2_{ajh}$.  The local shrinkage parameters $\lambda_{ajh}$ are assigned some prior $g_1$, allowing different amount of shrinkage on $B^*_{ajh}$ within the regularization subset $j$. The global shrinkage parameter $\psi_{aj}$ controls the overall level of shrinkage in the subset $j$, which leads to some type of smoothing over $t$ in $B_a(t)$, and is assigned a prior $g_2$ indexed by the hyperparameter $\Theta_{aj}$. 

Conditioning on $\psi_{aj}$ and integrating out $\lambda_{ajh}$, different choices of $g_1$ result in different marginal distributional forms that lead to different types of penalization and forms of regularization. A degenerate distribution $\lambda_{ajh} \sim \delta_{1}$ induces a Gaussian prior on $B^*_{ajh}$, leading to L$2$ penalization which would be a natural choice of regularization if spline basis functions are used. $\lambda_{ajh}^2 \sim \text{Exp}(\frac{1}{2})$ induces a Laplace prior on $B^*_{ajh}$, leading to L$1$ penalization and for which the maximum \textit{a posteriori} estimator is equivalent to the lasso estimate widely used for variable selection. $\lambda_{ajh} \sim C^+(0,1)$ induces a horseshoe prior~\citep{carvalho2009handling,carvalho2010horseshoe} on $B^*_{ajh}$, leading to non-linear adaptive shrinkage particularly desirable for wavelet transform, which tends to concentrate the signals in the data space on a relatively small number of wavelet coefficients that are usually large in magnitude, with the remaining coefficients being small and mostly consisting of noise. The infinitely tall spike of the horseshoe prior at the origin can strongly shrink the small coefficients, and its symmetric flat and Cauchy-like tails can prevent over-shrinkage of the large coefficients and retain the dominant local features in the observed data~\citep{carvalho2009handling}.

To summarize, our proposed model performs quantile regression on functional responses based on model \eqref{eq:1}, represents the coefficient functions using an appropriate basis representation as specified by model \eqref{eq:2}, and regularizes the basis coefficients by employing a global-shrinkage prior in model \eqref{eq:3}. Henceforth, we term this model as Bayesian functional quantile regression (FQR).

\medskip
\medskip

In practice, the functional responses are observed only on some discrete grid. Because our model is built for functional data sampled on a sufficiently dense grid, interpolation can be reasonably used to get a common grid for functional observations across subjects. If we assume that $\mathbf{Y}(t)=(Y_1(t),\dots, Y_N(t))'$ are all observed on a common grid $\mathbf{t}=(t_1, \dots, t_T)'$, and utilize the scale mixture representation of AL, we can represent the discrete version of model \eqref{eq:1} as \\[-25pt]
\begin{align} \label{eq:4}
Y_i(t_l)=\mathbf{X}'_i \mathbf{B}^{\tau}(t_l) + \frac{1-2\tau}{\tau(1-\tau)} \xi_i(t_l) + \sqrt{\frac{2\xi_i(t_l)\sigma(t_l)}{\tau(1-\tau)}} Z_i(t_l), 
\end{align}
for sample $i=1,\dots, N$ and position $l=1, \dots, T$.
In model \eqref{eq:4}, $\mathbf{Y}$ is an $N\times T$ matrix of functional responses with $Y_i(t_l)$ being the observation for sample $i$ at position $l$,  $\mathbf{B}$ is a $p\times T$ matrix of functional coefficients with its $l^{th}$ column  $\mathbf{B}^{\tau}(t_l)=(B^{\tau}_1(t_l), \dots, B^{\tau}_p(t_l))'$ being the vector of quantile regression coefficients at position $l$, $\sigma(t_l)$ is the scale parameter of the AL distribution at position $l$, $\xi_i(t_l)$ is the latent variable for sample $i$ at position $l$ following exponential distribution with mean $\sigma(t_l)$ independently across positions and samples, and $Z_i(t_l)$ is a standard normal variable $i.i.d$ across positions and samples.

Equation \eqref{eq:2} can now be expressed as \\[-20pt]
\begin{align} \label{eq:5}
\mathbf{B}=\mathbf{B}^*\mathbf{\Phi}, 
\end{align}
where $\mathbf{B}^*$ is a $p\times K$ matrix of basis coefficients,  $\mathbf{\Phi}$ is a full rank $K \times T$ matrix whose $k$th row corresponds to the basis function $\phi_k$ evaluated on the discrete grid $\mathbf{t}$.

\subsection{Posterior computation} \label{Gibbs}
We take a fully Bayesian approach to fit the FQR model. For appropriately chosen priors $g_1$ and $g_2$, posterior sampling proceeds via a scalable blocked Gibbs sampler with data augmentation if necessary. We outline the steps to draw posterior samples of the parameters in model \eqref{eq:4} as follows, and leave the full computational details in the supplementary materials.
\begin{enumerate}
   \item For each $l$, sample $(\sigma(t_l)|\mathbf{B}(t_l),\mathbf{y}(t_l))$ from an inverse Gamma distribution; \\[-20pt]
   \item For each $i$ and $l$, sample $(1/ \xi_i(t_l)|\mathbf{B}(t_l),\sigma(t_l),\mathbf{y}(t_l))$ from an inverse Gaussian distribution; \\[-20pt]
   \item For each $a$, sample $(\mathbf{B}^*_a|\mathbf{B}^*_{-a},\bm{\lambda}_a,\bm{\psi}_a,\bm{\xi},\bm{\sigma},\mathbf{Y})$ from multivariate normal; \\[-20pt]
   \item For each $a$, $j$, $h$, sample the local shrinkage parameter $(\lambda_{ajh}|B^*_{ajh},\psi_{aj})$; for each $a$, $j$, sample the global shrinkage parameter $(\psi_{aj}|\bm{\lambda}_{aj},\mathbf{B}^*_{aj})$; \\[-20pt]
   \item Project the rows of the updated basis coefficients $\mathbf{B}^*$ back to the data space using equation \eqref{eq:5}. 
\end{enumerate}

\subsection{Posterior inference} \label{inference}
The posterior samples obtained from the MCMC procedure can be used to construct a Bayesian estimator and perform Bayesian inference for any function of the parameters in model \eqref{eq:4}. In particular, for the functional coefficient $\mathbf{B}_a=(B_a(t_1),\dots,B_a(t_L))'$, a $100(1-\alpha)\%$ simultaneous credible band can be constructed from the posterior samples of $\mathbf{B}_a$ using the method described by~\citet{ruppert2003semiparametric} for $\alpha \in (0,1)$. Suppose $\{\mathbf{B}^{(g)}_a, g=1,\dots,G\}$ are the $G$ posterior samples of $\mathbf{B}_a$, where $\mathbf{B}^{(g)}_a=(B^{(g)}_a(t_1),\dots,B^{(g)}_a(t_T))'$. Let $m(B_a(t_l))$ and $\hat{sd}(B_a(t_l))$ denote the mean and standard deviation of $B_a(t_l)$ estimated from the $G$ posterior samples, a $100(1-\alpha)\%$ simultaneous credible band can be constructed by 
$$ \left[m(B_a(t_l))-q_{\alpha}\hat{sd}(B_a(t_l)),m(B_a(t_l))+q_{\alpha}\hat{sd}(B_a(t_l))\right], \quad l=1,\dots,T, $$
where $q_{\alpha}$ is the $(1-\alpha)$ sample quantile of
$$ \max\limits_{1 \leq l \leq T}\left|\frac{B_a^{(g)}(t_l)-m(B_a(t_l))}{\hat{sd}(B_a(t_l))}\right|,  \quad g=1,\dots,G. $$

\medskip

Given a quantile level $\tau$ and covariate $a$, it is often of interest to identify the locations $t$ for which $B_a(t)$ is significantly different from zero while accounting for multiple testing  in the functional data context. For example, in the pancreatic cancer mass spectrometry dataset, if the covariate $a$ denotes cancer status, then the identified locations $t$ would correspond to the spectral regions for which the $\tau$th quantile of protein expressions significantly differs between the cancer and normal populations. In this paper, we consider an approach that performs functional inference based on simultaneous band scores, or \textit{SimBaS}~\citep{meyer2015bayesian}, which involve inverting the joint credible bands for each $t$. SimBaS of a functional location $t_l$ is defined as the minimum $\alpha$ for which the $100(1-\alpha)\%$ simultaneous credible band excludes zero at $t_l$. At a pre-chosen level $\alpha$, we flag $t_l$ as significant if its SimBaS is less than or equal to $\alpha$. Given that it is based on the $100(1-\alpha)\%$ simultaneous credible band for which there is a $100(1-\alpha)\%$ posterior probability that the \textit{entire} function $B_a(t)$ lies within the corresponding band, use of this measure effectively accounts for multiple testing based on an experimentwise error rate like criterion.  

In terms of flagging significant spectral regions, the SimBaS account for statistical significance, but not practical significance. One may wish to also require a difference of some minimum effect size to flag a spectral region as significant, which can be specified as a minimum fold change $\delta$ if the log spectral intensities are measured.  In that case, one may require SimBaS$<\alpha$ and 
 $\left|B_a(t)\right|\geq\log_2\delta$, requiring at least a $\delta$-fold change for the $\tau$th quantile of protein expressions between cancer and normal groups, quantified by posterior mean estimates of $B_a(t)$. 

\subsection{Sandwich likelihood correction} \label{correction}
We note that the AL likelihood is used as a working likelihood in our Bayesian framework, which is not likely to be the true data generating likelihood. Recent studies raised concerns about the validity of posterior inference based on the AL working likelihood \citep{yang2016posterior, sriram2015sandwich, syring2018calibrating}. More specifically, for any given location $t$, when assigned a proper prior, the posterior distribution of the $p \times 1$ vector $\mathbf{B}^{\tau}(t)$ is shown to be approximately normal centered at $\tilde{\mathbf{B}}^{\tau}(t) = m(\mathbf{B}^{\tau}(t))$ for large $n$, but its scaled posterior covariance matrix $n \:\!\tilde{\Sigma}^{\tau}(t)$ does not converge to the asymptotic covariance of $n^{1/2} \:\! \hat{\mathbf{B}}^{\tau}(t)$ as established in \citet{koenker2005quantile}, where $\hat{\mathbf{B}}^{\tau}(t)$ is the M-estimator of $\mathbf{B}^{\tau}(t)$ by minimizing the check loss function. This suggests that the $100 \:\! (1-\alpha)\%$ Bayesian credible sets based on the AL likelihood in general do not have a frequentist coverage of $1-\alpha$. These studies also proposed simple adjustment strategies to achieve asymptotically valid posterior inference. Among them, \citet{sriram2015sandwich} showed that if assume any fixed scale parameter $\sigma(t)$ and construct a ``sandwich likelihood" specified in \eqref{eq:6}, \\[-20pt]
\begin{align} \label{eq:6}
\quad\quad\quad p\left(\mathcal{D}(t) \:|\: \mathbf{B}^{\tau}(t) \right) \propto \exp \left[-\frac{1}{2} \left(\tilde{\mathbf{B}}^{\tau}(t) \!-\! \mathbf{B}^{\tau}(t) \right)' \tilde{\Sigma}^{\tau}_{\text{adj}}(t)^{-1} \! \left(\tilde{\mathbf{B}}^{\tau}(t) \!-\! \mathbf{B}^{\tau}(t) \right) \right], \\[-25pt]
\end{align}
where $\mathcal{D}(t)$ represents the observed data at $t$, $\tilde{\Sigma}^{\tau}_{\text{adj}}(t) = \frac{n \tau (1-\tau)}{\sigma^2(t)} \tilde{\Sigma}^{\tau}(t) \tilde{D}_0 \tilde{\Sigma}^{\tau}(t)$ and $\tilde{D}_0 = n^{-1} \: \mathbf{X}'\mathbf{X}$, then the Bayesian credible sets of $\mathbf{B}^{\tau}(t)$ based on this sandwich likelihood and a proper prior have the nominal frequentist coverage asymptotically. 

Motivated by these concerns, we also considered an adjusted version of our Bayesian FQR model to improve the frequentist properties of posterior inference in the simulation studies and real data application, in which we replace the AL likelihood with the Gaussian sandwich likelihood in \eqref{eq:6} at each location $t_l \; (l=1, \dots, T)$. Since the adjusted posterior covariance $\tilde{\Sigma}^{\tau}_{\text{adj}}(t)$ is shown to be asymptotically invariant in the value of the scale parameter $\sigma(t)$~\citep{yang2016posterior}, we fix $\sigma(t) = 1$ at each $t$ for convenience. The posterior sampling of the adjusted Bayesian FQR proceeds in a similar manner as the Bayesian FQR, and the full computational details are provided in the supplementary materials.

\section{Simulation studies} \label{simulations}
We conducted simulation studies to evaluate the performance of our proposed model and compare to several straightforward approaches that people might use in the FQR setting. 

\textbf{Simulation design:} 
The shapes of mass spectrometry peaks can be approximated by Gaussian densities~\citep{zhang2009review}, with the heights of the peaks roughly quantifying the relative abundance of proteins at the corresponding spectral locations. Thus, in constructing a simulation to mimic mass spectrometry data, we utilize peaks with Gaussian shapes. Specifically, functional data were generated based on the following model,  \\[-20pt]
\begin{align} \label{eq:7}
\begin{split}
\mathbf{y}_i(t) =& \sum\limits_{k=1}^{4}c_{i,k}\varphi\left(t\mid\mu_{k},\sigma_{k}\right)+ \mathbf{e}_i(t), \\
c_{i,k} =& \mathbf{1}\{x_{i2}=-1\} f_{1,k} + \mathbf{1}\{x_{i2}=1\} f_{2,k} + x_{i3} \alpha_k,  
\end{split}
\end{align}
with a sample size of $N=400$ subjects indexed by $i$, and $K=4$ non-overlapping peaks indexed by $k$. $\varphi\left(t\mid\mu_{k},\sigma_{k}\right)$ is the probability density function of a normal distribution with mean $\mu_{k}$ and standard deviation $\sigma_{k}$, which corresponds to a Gaussian shaped peak in $\mathbf{y}(t)$ centered at $\mu_{k}$. The design matrix $\mathbf{X}$ consists of 3 columns: an intercept $x_1$, a binary variable $x_2$ taking values from $\{-1,1\}$ with equal probability, and an independent standard normal variable $x_3$. In the context of mass spectrometry data, $x_2$ can be interpreted as a group indicator of each subject, i.e., whether the subject belongs to the cancer cohort or the normal cohort. $x_3$ can be interpreted as a continuous demographic or clinical factor that is rescaled to have a standard normal distribution in the population and is potentially predictive of expression levels of certain proteins. $c_{i,k}$, which is determined jointly by $x_{i2}$ and $x_{i3}$, dictates the magnitude of peak $k$ in the funcional observation $\mathbf{y}_i(t)$. $\mathbf{e}(t)$, the noise term assumed to be $i.i.d$ across subjects, is a $\text{Gaussian AR}(1)$ process with lag $1$ autocorrelation $\rho=0.5$ and a marginal distribution $e(t) \sim N(0,9)$. The functional response $\mathbf{y}(t)$ is observed on an equally spaced grid of $301$ on the interval $\left[0,9\right]$. The distributions of $f_{1,k},\; f_{2,k}$ and the values taken by $\mu_{k}$, $\sigma_{k}$ and $\alpha_k$ are provided in Table \ref{table:1}. It should be noted that while the noise term $\mathbf{e}(t)$ in our simulation setup \eqref{eq:7} is Gaussian, the conditional distribution $p(\mathbf{y}(t)|\mathbf{x})$ in many cases is not Gaussian. This is because the curve-to-curve variations include both the residual terms $\mathbf{e}(t)$ and the stochastic functional components induced by $c_{i,k}$ in \eqref{eq:7}, which in turn depend on $f_{1,k}$ or $f_{2,k}$. A non-Gaussian distribution of $f_{1,k}$ or $f_{2,k}$, such as $t_2$ or inverse Gamma presented in Table \ref{table:1}, induces curve-to-curve deviations that are not Gaussian.  We simulated $100$ replicate datasets.

\begin{table}[h!]
\caption{\textbf{\textit{Parameter specifications of the data generating models in simulations.}}}
\label{table:1}
\setlength{\tabcolsep}{16pt}
\begin{tabular*}{\linewidth}{ @{} *{12}c @{}}
\toprule
Basis index $k$ & $\mu_{k}$ & $\sigma_{k}$ & $f_{1,k}$ & $f_{2,k}$ & $\alpha_k$ \\[3pt]
\midrule
1 & 1 & 0.18 & $1.75t_{2}+30$ & $N(30,1^2)$ & 0 \\[2pt]
2 & 3.25 & 0.18 & $N(30,1^2)$ & $N(30,1^2)$ & 0 \\[2pt]
3 & 5.5 & 0.18 & $N(30.5,0.4^2)$ & $IG(1,0.35)+30$ & 0 \\[2pt]
4 & 8 & 0.18 & $N(30,1^2)$ & $N(30,1^2)$ & 1 \\
\bottomrule
\end{tabular*}
\end{table}

At a given quantile $\tau$, the model $\mathbf{Y}=\mathbf{X}\mathbf{B}^{\tau}+\mathbf{E}^{\tau}$ is fitted to perform FQR, with $\mathbf{Y}$ being the $400 \times 301$ functional response matrix and $\mathbf{X}$ being the $400 \times 3$ design matrix. The quantities of interest are: (1) the group effect function $B_2^{\tau}(t)$, which quantifies the difference in the $\tau${th} quantile at position $t$ between the two groups indexed by $x_2$ while conditioning on $x_3$, and (2) $B_3^{\tau}(t)$, which quantifies the change in peak heights if the continuous predictor $x_3$ increases by one unit while conditioning on $x_2$.   

\begin{figure}[h]
    \centering
    \includegraphics[width=0.5\textwidth]{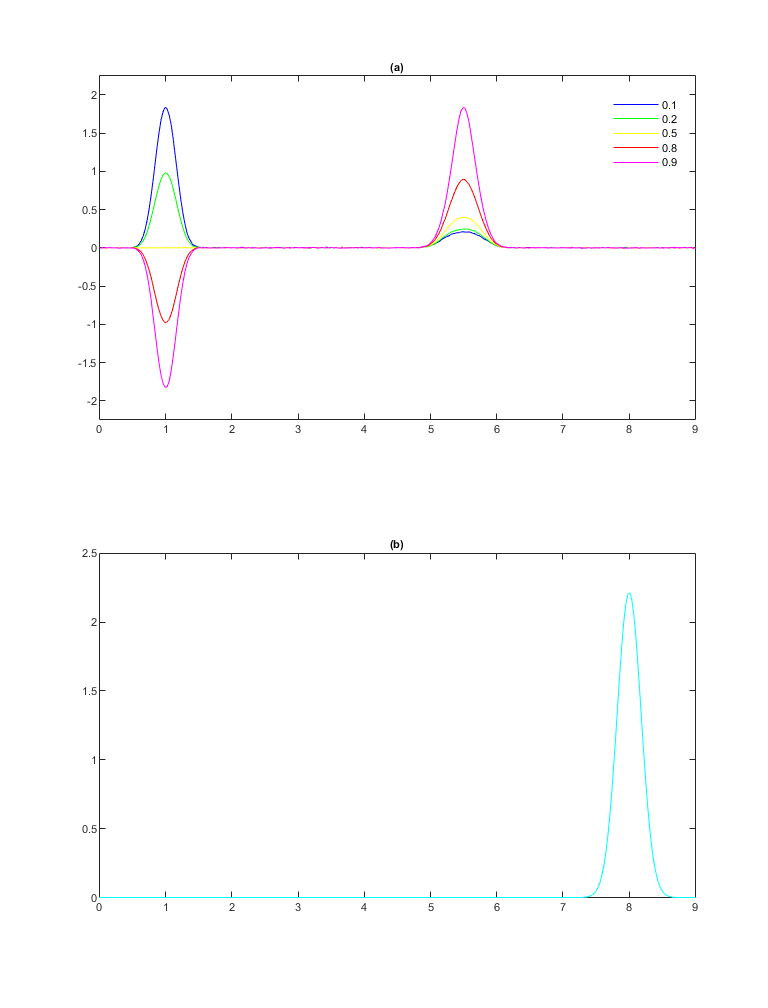}
    \vspace{-6mm}
   \caption{\textbf{\textit{Ground truth for functional coefficients of interest.}} The true group effect functions $B_2^{\tau}(t)$ at multiple quantile levels are shown in (a). The true $B_3^{\tau}(t)$, which quantifies the change in peak magnitudes per unit change in $x_3$ conditional on other covariates and is constant across different quantile levels, is shown in (b). }  
    \label{fig:figure3}
    \vspace{-2mm}
\end{figure}

The true group effect functions $B_2^{\tau}(t)$ at various levels of $\tau$ are shown in Figure~\ref{fig:figure3} (a). Conditional on $x_3$, obvious group differences are present at $\tau=0.1, 0.2, 0.8, 0.9$ at the first peak, and at $\tau=0.8, 0.9$ at the third peak. For the first peak, these group differences would not be detected by mean or median regression on the simulated data, because the magnitudes associated with the first peak are purposely designed to have identical mean and median between two groups when conditioning on $x_3$, but the symmetric heavy tailed $t_2$ chosen for $f_{1,1}$ leads to remarkable group differences at more extreme quantiles. For the third peak, the choice of an inverse Gamma distribution without a finite mean for $f_{2,3}$ renders it theoretically implausible to perform mean regression on the simulated data, while its heavily right skewed nature makes the distributions of the simulated spectral intensities at the third peak greatly differ in the upper tail but not the median or lower tail between two groups when conditioning on $x_3$. This design is motivated by the setting whereby group differences are evident in the tails but not the center of the distribution, which we observed from Figure~\ref{fig:figure2} in the spectral region ($5700$D, $6000$D) and described in Section \ref{motivation}, and allows us to examine the performance of our proposed approach in different types of heavy tailed settings. 

The true functional coefficient $B_3^{\tau}(t)$, which is constant across different quantile levels, is shown in Figure~\ref{fig:figure3} (b). Conditioning on $x_2$, $\alpha_k$ represents the change in the magnitude of peak $k$ that is caused by one unit increase of the continuous covariate $x_3$.

\textbf{Bayesian FQR model:}  We applied our Bayesian FQR model to these simulated data, using a wavelet basis with a Daubechies wavelet with $4$ vanishing moments, periodic boundary conditions, and a decomposition level $J=6$, and a horseshoe regularization prior. Note that we did not simulate data with AL residual errors, nor were wavelets used in any way in simulating the data. Therefore, the data generating process for the simulated data does not give any inherent advantage to our approach over others.

\textbf{Alternative approaches:} In addition to our proposed Bayesian FQR approach, we also considered a few alternative approaches and assessed their performance, including $1)$ the na\"ive Bayesian quantile regression, or Bayesian QR~\citep{yu2001bayesian} which performs Bayesian quantile regression separately at each location $t$ using the AL likelihood. $2)$ the adjusted Bayesian FQR as proposed in \ref{correction}. $3)$ the na\"ive quantile regression, or QR~\citep{koenker2005quantile} which does quantile regression at each individual location $t$ by minimizing the check loss function. $4)$ QR with spline smoothing, which smooths the functional coefficients estimated by QR using splines. All unique values of $t$ are used as knots to determine the spline basis functions, and the smoothing parameter is chosen by generalized cross-validation. $5)$ QR with wavelet denoising, which denoises the functional coefficients estimated by QR by projecting them into the wavelet domain and placing minimax hard thresholding on the wavelet coefficients. $6)$ FDboost, which fits a functional linear array model by component-wise gradient boosting.

It should be pointed out that the two-step methods 4) and 5), while perhaps natural ideas to consider, have not to our knowledge been used in the literature to perform FQR, so are in a sense themselves new methods introduced in this paper, but we hypothesize that our unified approach will have inferential advantages over them. We implemented the Bayesian approaches in MATLAB \citep{MATLAB2016} and ran each MCMC chain for $8000$ iterations, discarding the first $2000$ and keeping every $3$. For the approaches 3)-5), we called the ``quantreg" package~\citep{quantreg2017} in R \citep{R2017} to do quantile regression, and performed bootstrap on the entire functional response $\mathbf{y}(t)$ and the covariate $\mathbf{x}$ to do inference. $2000$ bootstrap samples were generated per case. We called the ``FDboost" package~\citep{fdboost2017} in R to implement approach $6$). 

\textbf{Evaluation criteria:} At each of the quantile levels $0.1, 0.2, 0.5, 0.8, 0.9$, Bayesian FQR model and alternative methods were applied to the simulated datasets to perform FQR. We used SimBaS to identify regions of the functional coefficients $B_a^{\tau}(t)\;(a=2, 3)$ where the absolute magnitude exceeds some practically meaningful threshold $\delta$ at each quantile level. Given the true $B_2^{\tau}(t)$ and $B_3^{\tau}(t)$ in this simulation, we chose $\delta = 0.3$ here. For non-Bayesian approaches, bootstrap samples were used in place of posterior samples to construct simultaneous confidence bands and compute SimBaS. At a given level $\alpha$, we flagged a location $t$ as significant if the SimBaS at $t$ is less than or equal to $\alpha$, and computed the sensitivity and false positive rate for detecting sites of at least size $\delta = 0.3$ for each approach. 

We also evaluated the estimation performance of these methods using (i) the integrated mean squared error (IMSE), (ii) the coverage probability of $95\%$ simultaneous band covering the true values, and (iii) the average width of $95\%$ simultaneous band across $t$. For a functional parameter $\theta(t)\;(t\in T)$ with true value $\theta_{0}(t)$, suppose $\left\{\hat{\theta}^{(m)}(t),\;m=1,\dots, M\right\}$ are the mean estimates computed from $M$ replicate datasets. For a replication $m$, IMSE is defined as $\int_{T}\left\{\hat{\theta}^{(m)}(t)-\theta_{0}(t)\right\}^2 dt$. 

\textbf{Simulation results:} Table \ref{table:2} summarizes the estimation and inferential performance of these methods at each quantile for $B_2^{\tau}(t)$ (upper table) and $B_3^{\tau}(t)$ (lower table). Where applicable, these summary measures are averaged over $100$ replicate datasets with standard deviations in parentheses.

The total time to perform FQR on a simulated dataset at the $5$ quantile levels on a $64$-bit operating system with 2 processors and an RAM of $32$GB was about $40$ minutes for Bayesian QR, $75$ minutes for Bayesian FQR with or without adjustment, and $60$ minutes for the bootstrap-based approaches with or without smoothing. This indicates that the Bayesian FQR is computationally scalable to high-dimensional functional datasets and on the same order of magnitude as the potential competing approaches.

At each quantile level $\tau$ considered, the Bayesian FQR and the adjusted Bayesian FQR clearly outperformed the na\"ive Bayesian QR by having better estimation accuracy (IMSE) and lower posterior variability, which is reflected by the narrower credible bands, for both $B_2^{\tau}(t)$ and $B_3^{\tau}(t)$. They also had substantially increased sensitivity for detection of significant regions in $B_2^{\tau}(t)$ at each of the commonly used levels $\alpha$, compared to na\"ive Bayesian QR. The same conclusions applied to the comparison between the bootstrap-based QR with spline smoothing and its na\"ive counterpart. These comparisons indicate that proper regularization of the functional coefficients leads to greatly improved performance in both estimation and inference.

Compared to the bootstrap-based methods with smoothing, the Bayesian FQR and the adjusted Bayesian FQR had similar or better estimation accuracy in all cases; in terms of inference, both of them had much tighter simultaneous credible band with similar coverage, and considerably higher sensitivity for detecting significant functional regions in $B_2^{\tau}(t)$ than the bootstrap-based methods with smoothing. Note that at each commonly used threshold $\alpha$, all the bootstrap-based methods have a very low sensitivity ($<0.3$) for discovery of significant sites in $B_2^{\tau}(t)$ at each quantile level considered. 

Comparing the Bayesian FQR with and without adjustment, the sandwich likelihood correction led to improved estimation accuracy, slightly wider simultaneous credible band and marginally higher coverage in all cases. In terms of detection of significant regions, the false positive rates of the original Bayesian FQR are already negligibly small; the adjustment further reduced the false positive rate to 0 in almost all cases, which is accompanied with a decrease in the sensitivity that is more pronounced for $B_2^{\tau}(t)$.

We also applied FDboost to our simulated data, but found that it did not appear to be suitable for these spiky, spatially heterogeneous functional data, and did not scale up well to the densely sampled data as considered in our simulations. Details about our implementation of FDboost are provided in the supplement.

\clearpage

\begin{landscape}
\captionsetup{font=footnotesize}
\begin{table}
\scriptsize
\caption{\textbf{\textit{Simulation results.}} For the Bayesian FQR and alternative methods, the sensitivity ($\times10^{-2}$) and false positive rate ($\times10^{-2}$) for detecting functional regions of at least size $\delta=0.3$ based on SimBaS at commonly used levels of $\alpha$, as well as the integrated mean squared error (IMSE), the coverage probability and average width of $95\%$ simultaneous band are presented for $B_2^{\tau}(t)$ in the upper table, and $B_3^{\tau}(t)$ in the lower table. Standard deviations over 100 replicates are given in parentheses where applicable. QR (+s) and QR (+w) refer to the bootstrap-based two-step approaches with spline smoothing and wavelet denoising respectively.\vspace*{-5pt}}
\label{table:2}
\begin{tabular*}{\linewidth}{ @{\extracolsep{\fill}}  *{15}c @{}} 
\toprule
$\tau$ & Methods & \multicolumn{4}{c}{Sensitivity ($\times10^{-2}$)} &
\multicolumn{4}{c}{False Positive Rate ($\times10^{-2}$)}  &  IMSE  & \makecell{Coverage \\ Joint Band} & \makecell{Ave Width \\ Joint Band}   \\
\midrule \midrule \vspace*{-12pt} \\ 
& $\alpha$ & 0.001 & 0.01 & 0.05 & 0.10  & 0.001 & 0.01  & 0.05 & 0.10 & \\
\cmidrule{3-6} \cmidrule{7-10} \cmidrule{11-11} \cmidrule{12-13} \vspace*{-11pt} \\
\multirow{5}{*}{$0.1$} & Bayes QR & 42.8 & 53.7 & 61.5 & 64.4 & 0.5 & 1.2 & 2.4 & 3.2 & 21.1(3.1) & 0.977 & 1.24(0.02)\\[-4pt]
                       & Bayes FQR & 64.1 & 72.8 & 78.5 & 81.3 & 0 & 0.1 & 0.4 & 0.7 & 9.8(2.8) & 0.992 & 0.97(0.03)\\[-4pt]
                       & Bayes FQR (+adj) & 23.5 & 46.3 & 64.8 & 72.2 & 0 & 0 & 0 & 0 & 5.9(2.8) & 0.998 & 1.11(0.05)\\[-4pt]
                       & QR & 0.3 & 1.6 & 4.6 & 8.1 & 0 & 0 & 0 & 0 & 19.8(3.0) & $>0.999$ & 2.42(0.04)\\[-4pt]
                       & QR (+s) & 0.3 & 5.4 & 19.0 & 29.9 & 0 & 0 & 0 & 0 & 6.9(2.8) & $>0.999$ & 1.27(0.07)\\[-4pt]
                       & QR (+w) & 0 & 0 & 0.2 & 0.6 & 0 & 0 & 0 & 0 & 8.2(2.9) & $>0.999$ & 2.45(0.05)\\[-2pt]
                       \midrule \vspace*{-13pt}  \\
\multirow{5}{*}{$0.2$} & Bayes QR & 10.1 & 19.6 & 27.6 & 32.4 & 0.1 & 0.3 & 0.7 & 1.0 & 14.3(1.8) & 0.993 & 1.23(0.02)\\[-4pt]
                       & Bayes FQR & 31.3 & 50.1 & 66.5 & 73.3 & 0 & 0 & 0.1 & 0.3 & 5.1(1.6) & 0.997 & 0.83(0.04)\\[-4pt]
                       & Bayes FQR (+adj) & 7.7 & 25.3 & 44.3 & 53.4 & 0 & 0 & 0 & 0 & 3.8(1.6) & 0.998 & 0.91(0.04)\\[-4pt] 
                       & QR & 0.4 & 0.8 & 3.3 & 5.1 & 0 & 0 & 0 & 0 & 13.7(1.7) & $>0.999$ & 1.90(0.02)\\[-4pt]
                       & QR (+s) & 0.3 & 1.9 & 10.0 & 16.1 & 0 & 0 & 0 & 0 & 5.0(1.6) & $>0.999$ & 1.09(0.06)\\[-4pt]
                       & QR (+w) & 0 & 0 & 0.1 & 0.4 & 0 & 0 & 0 & 0 & 5.8(1.5) & $>0.999$ & 1.95(0.03)\\[-2pt]
                       \midrule \vspace*{-13pt} \\
\multirow{5}{*}{$0.5$} & Bayes QR & - & - & - & - & 0 & 0.1 & 0.2 & 0.3 & 11.0(1.2) & 0.998 & 1.23(0.03)\\[-4pt]
                       & Bayes FQR & - & - & - & - & 0 & 0 & 0.1 & 0.1 & 2.8(1.0) & 0.999 & 0.73(0.03)\\[-4pt]
                       & Bayes FQR (+adj) & - & - & - & - & 0 & 0 & 0 & 0 & 2.3(0.9) & $>0.999$ & 0.79(0.03)\\[-4pt] 
                       & QR & - & - & - & - & 0 & 0 & 0 & 0 & 10.7(1.2) & $>0.999$ & 1.62(0.02)\\[-4pt]
                       & QR (+s) & - & - & - & - & 0 & 0 & 0 & 0 & 4.2(1.2) & $>0.999$ & 0.99(0.04)\\[-4pt]
                       & QR (+w) & - & - & - & - & 0 & 0 & 0 & 0 & 4.7(1.1) & $>0.999$ & 1.66(0.02)\\[-2pt]
                       \midrule \vspace*{-13pt} \\
\multirow{5}{*}{$0.8$} & Bayes QR & 5.8 & 11.9 & 18.6 & 23.2 & 0.1 & 0.3 & 0.7 & 1.0 & 15.0(2.4) & 0.993 & 1.28(0.07)\\[-4pt]
                       & Bayesian FQR & 28.5 & 49.5 & 64.1 & 70.1 & 0 & 0.1 & 0.2 & 0.3 & 5.8(2.1) & 0.995 & 0.84(0.04)\\[-4pt]
                       & Bayes FQR (+adj) & 16.5 & 37.4 & 56.2 & 65.1 & 0 & 0 & 0 & 0 & 4.6(2.0) & 0.997 & 0.93(0.04)\\[-4pt] 
                       & QR & 0.2 & 0.9 & 2.9 & 5.0 & 0 & 0 & 0 & 0 & 14.5(2.3) & $>0.999$ & 1.95(0.03)\\[-4pt]
                       & QR (+s) & 0.4 & 3.9 & 15.7 & 25.9 & 0 & 0 & 0 & 0 & 5.7(2.1) & $>0.999$ & 1.13(0.06)\\[-4pt]
                       & QR (+w) & 0 & 0 & 0.2 & 0.5 & 0 & 0 & 0 & 0 & 6.6(2.0) & $>0.999$ & 1.99(0.04)\\[-2pt]
                       \midrule \vspace*{-13pt}  \\
\multirow{5}{*}{$0.9$} & Bayes QR & 26.1 & 36.2 & 45.9 & 49.8 & 0.5 & 1.2 & 2.4 & 3.1 & 26.5(8.9) & 0.978 & 1.38(0.14)\\[-4pt]
                       & Bayes FQR & 57.1 & 69.3 & 76.9 & 79.5 & 0.1 & 0.2 & 0.6 & 0.9 & 13.5(5.6) & 0.986 & 1.02(0.04)\\[-4pt]
                       & Bayes FQR (+adj) & 27.8 & 53.4 & 70.1 & 75.9 & 0 & 0 & 0 & 0 & 9.5(4.2) & 0.995 & 1.16(0.06)\\[-4pt] 
                       & QR & 0 & 0.2 & 1.7 & 3.5 & 0 & 0 & 0 & 0 & 25.8(10.7) & $>0.999$ & 2.70(0.15)\\[-4pt]
                       & QR (+s) & 0.1 & 1.1 & 7.7 & 16.0 & 0 & 0 & 0 & 0 & 12.5(10.2) & $>0.999$ & 1.52(0.16)\\[-4pt]
                       & QR (+w) & 0 & 0 & 0 & 0.1 & 0 & 0 & 0 & 0 & 14.4(10.4) & $>0.999$ & 2.75(0.17)\\[-2pt]
\bottomrule
\end{tabular*}
\end{table}

\pagebreak

\begin{table}
\scriptsize
\begin{tabular*}{\linewidth}{ @{\extracolsep{\fill}}  *{15}c @{}} 
\toprule
$\tau$ & Methods & \multicolumn{4}{c}{Sensitivity ($\times10^{-2}$)} &
\multicolumn{4}{c}{False Positive Rate ($\times10^{-2}$)}  &  IMSE  & \makecell{Coverage \\ Joint Band} & \makecell{Ave Width \\ Joint Band}   \\
\midrule \midrule \vspace*{-12pt} \\ 
& $\alpha$ & 0.001 & 0.01 & 0.05 & 0.10  & 0.001 & 0.01  & 0.05 & 0.10 & \\
\cmidrule{3-6} \cmidrule{7-10} \cmidrule{11-11} \cmidrule{12-13} \vspace*{-11pt} \\
\multirow{5}{*}{$0.1$} & Bayes QR & 72.7 & 78.5 & 81.5 & 83.0 & 0.4 & 1.1 & 2.1 & 2.8 & 20.0(2.4) & 0.979 & 1.26(0.04)\\[-3pt]
                       & Bayes FQR & 78.5 & 83.8 & 87.2 & 88.8 & 0.1 & 0.2 & 0.7 & 1.1 & 8.9(1.9) & 0.992 & 0.96(0.04)\\[-3pt]
                       & Bayes FQR (+adj) & 62.1 & 72.7 & 79.6 & 82.8 & 0 & 0 & 0 & 0 & 5.2(1.6) & $>0.999$ & 1.11(0.07)\\[-3pt] 
                       & QR & 22.3 & 37.8 & 49.0 & 53.4 & 0 & 0 & 0 & 0 & 18.6(2.3) & $>0.999$ & 2.53(0.09)\\[-3pt]
                       & QR (+s) & 69.1 & 76.8 & 83.4 & 85.6 & 0 & 0 & 0 & 0 & 6.2(1.7) & $>0.999$ & 1.26(0.09)\\[-3pt]
                       & QR (+w) & 16.1 & 34.3 & 47.3 & 52.1 & 0 & 0 & 0 & 0 & 7.5(1.6) & $>0.999$ & 2.43(0.09) \\[-1pt]
                       \midrule \vspace*{-11pt} \\
\multirow{5}{*}{$0.2$} & Bayes QR & 73.0 & 78.3 & 82.3 & 83.7 & 0.1 & 0.3 & 0.6 & 1.0 & 13.8(1.6) & 0.994 & 1.24(0.04)\\[-3pt]
                       & Bayes FQR & 81.4 & 87.7 & 91.3 & 93.2 & 0 & 0 & 0.1 & 0.3 & 4.8(1.1) & 0.998 & 0.83(0.04)\\[-3pt]
                       & Bayes FQR (+adj) & 71.0 & 79.8 & 87.0 & 90.5 & 0 & 0 & 0 & 0 & 3.9(1.1) & $>0.999$ & 0.92(0.04)\\[-3pt] 
                       & QR & 42.8 & 55.4 & 62.7 & 65.5 & 0 & 0 & 0 & 0 & 13.3(1.6) & $>0.999$ & 2.02(0.07)\\[-3pt]
                       & QR (+s) & 75.3 & 81.3 & 86.8 & 88.8 & 0 & 0 & 0 & 0 & 5.0(1.3) & $>0.999$ & 1.10(0.07)\\[-3pt]
                       & QR (+w) & 36.3 & 51.4 & 59.8 & 64.0 & 0 & 0 & 0 & 0 & 5.7(1.2) & $>0.999$ & 1.97(0.07)\\[-1pt]
                       \midrule \vspace*{-11pt} \\
\multirow{5}{*}{$0.5$} & Bayes QR & 74.5 & 79.4  & 82.7 & 83.8 & 0 & 0.1 & 0.2 & 0.3 & 10.8(1.5) & 0.998 & 1.24(0.05)\\[-3pt]
                       & Bayes FQR & 82.9 & 89.3 & 94.0 & 95.6 & 0 & 0 & 0.1 & 0.3 & 3.5(1.0) & 0.998 & 0.74(0.03)\\[-3pt]
                       & Bayes FQR (+adj) & 76.4 & 85.0 & 90.8 & 93.3 & 0 & 0 & 0 & 0.1 & 3.3(1.0) & $>0.999$ & 0.80(0.04)\\[-3pt] 
                       & QR & 54.5 & 62.5 & 69.4 & 72.2 & 0 & 0 & 0 & 0 & 10.6(1.5) & $>0.999$ & 1.72(0.06)\\[-3pt]
                       & QR (+s) & 78.8 & 84.2 & 87.7 & 89.5 & 0 & 0 & 0 & 0 & 4.6(1.2) & $>0.999$ & 1.01(0.06)\\[-3pt]
                       & QR (+w) & 49.5 & 58.3 & 66.0 & 70.0 & 0 & 0 & 0 & 0 & 4.9(1.1) & $>0.999$ & 1.69(0.06)\\[-1pt]
                       \midrule \vspace*{-11pt} \\
\multirow{5}{*}{$0.8$} & Bayes QR & 74.0 & 79.0 & 82.5 & 83.7 & 0.1 & 0.3 & 0.6 & 1.0 & 14.1(1.8) & 0.994 & 1.29(0.07)\\[-3pt]
                       & Bayesian FQR & 80.7 & 87.5 & 92.1 & 94.0 & 0 & 0.1 & 0.3 & 0.4 & 4.9(1.3) & 0.997 & 0.84(0.05)\\[-3pt]
                       & Bayes FQR (+adj) & 71.9 & 79.8 & 86.3 & 90.1 & 0 & 0 & 0 & 0.1 & 4.0(1.2) & $>0.999$ & 0.92(0.05)\\[-3pt] 
                       & QR & 43.8 & 54.2 & 62.4 & 65.6 & 0 & 0 & 0 & 0 & 13.6(1.8) & $>0.999$ & 2.05(0.07)\\[-3pt]
                       & QR (+s) & 75.3 & 81.3 & 86.5 & 88.9 & 0 & 0 & 0 & 0 & 5.3(1.5) & $>0.999$ & 1.13(0.08)\\[-3pt]
                       & QR (+w) & 36.5 & 50.9 & 60.1 & 62.9 & 0 & 0 & 0 & 0 & 5.9(1.4) & $>0.999$ & 2.00(0.08)\\[-1pt]
                       \midrule \vspace*{-11pt} \\
\multirow{5}{*}{$0.9$} & Bayes QR & 72.8 & 78.0 & 81.4 & 83.1 & 0.4 & 1.0 & 1.9 & 2.6 & 21.3(3.4) & 0.980 & 1.36(0.11)\\[-3pt]
                       & Bayes FQR & 76.8 & 83.9 & 88.5 & 90.3 & 0 & 0.2 & 0.6 & 1.1 & 9.3(2.7) & 0.993 & 1.00(0.05)\\[-3pt]
                       & Bayes FQR (+adj) & 63.5 & 73.4 & 81.4 & 85.2 & 0 & 0 & 0 & 0.1 & 5.8(2.2) & $>0.999$ & 1.12(0.07)\\[-3pt] 
                       & QR & 21.7 & 37.3 & 49.8 & 54.5 & 0 & 0 & 0 & 0 & 20.1(3.3) & $>0.999$ & 2.63(0.10)\\[-3pt]
                       & QR (+s) & 69.5 & 77.3 & 83.8 & 87.1 & 0 & 0 & 0 & 0 & 7.2(2.9) & $>0.999$ & 1.32(0.10)\\[-3pt]
                       & QR (+w) & 15.8 & 33.3 & 46.0 & 52.1 & 0 & 0 & 0 & 0 & 8.7(2.7) & $>0.999$ & 2.53(0.10)\\
\bottomrule
\end{tabular*}

\end{table}
\end{landscape}

\clearpage

\section{Functional Quantile Regression for Protein Biomarker Discovery} \label{application}

We applied our Bayesian FQR model using wavelet basis functions, as well as the alternative methods described in Section \ref{simulations} to perform FQR on the pancreatic cancer mass spectrometry dataset at
$\tau=0.1, 0.25, 0.5, 0.75, 0.9$. We are primarily interested in identifying regions of the mass spectra that significantly differ between the cancer and normal group at each quantile level while accounting for multiple testing, and comparing the flagged regions across different quantiles. For comparative purpose, we also applied the wavelet-based functional mixed model, or WFMM
~\citep{morris2006wavelet} to perform functional mean regression to assess which results found by the Bayesian FQR would have been missed had only functional mean regression been done.

Our analysis is focused on the part of the spectra from $t=5,000$ to $t=8,000$ Daltons including $1,659$ observations per spectrum.  To draw meaningful biological conclusions from the mass spectrometry data, it is critical to perform appropriate preprocessing before further statistical analysis~\citep{sorace2003data}. The preprocessing steps for MALDI-TOF mass spectrometry data include baseline correction, normalization and denoising, which were performed using the methods described by~\citet{coombes2005improved}. The spectral intensities can span several orders of magnitude across mass-to-charge ratio $t$ for a given sample, and demonstrate extreme skewness across samples at a given $t$. To mitigate these issues, we took $\log_2$ transformation on the mass spectrometry data, which also allows an absolute difference of one on the $\log_2$ scale to be interpreted as a two-fold change on the original scale. These samples were processed in four different blocks over a span of several months.  Previous studies~\citep{baggerly2003comprehensive,baggerly2004reproducibility} show that block effects associated with MALDI-TOF instruments can often be severe, so we estimated and subtracted the block-specific mean from the preprocessed mass spectra to adjust for the block effects. In Figure~\ref{fig:figure1}, the right column displays the corresponding preprocessed spectra of the raw spectra in the left column, and this comparison clearly shows the effect of preprocessing.

The design matrix $\mathbf{X}$ for this dataset is a $256 \times 2$ matrix, with the first column being the intercept and the second column denoting cancer (=$1$) or normal (=$-1$) status. The models $\mathbf{Y}=\mathbf{X}\mathbf{B}^{\tau}+\mathbf{E}^{\tau}(\tau=0.1, 0.25, 0.5, 0.75, 0.9)$ and $\mathbf{Y}=\mathbf{X}\mathbf{B}^{\text{mean}}+\mathbf{E}^{\text{mean}}$ are individually fitted to perform FQR and functional mean regression. The cancer main effect functions $B_2^{\tau}(t)$ and $B_2^{\text{mean}}(t)$ respectively quantify the difference in the $\tau${th} quantile and mean of the $\log_2$ spectral intensities between cancer and normal groups at the spectral location $t$. For the Bayesian FQR model with and without adjustment, we performed discrete wavelet transform (DWT) using the Daubechies wavelet with $4$ vanishing moments, periodic boundary conditions, and a decomposition level $J=8$. We placed a horseshoe prior on  $B^*_{ajh}$, assuming $\lambda_{ajh} \sim C^+(0,1)$ and $\psi_{aj} \sim C^+(0,s_a)$, where $s_a$ is a hyperparameter with a vague hyperprior $s_a^2 \sim \text{inverse Gamma} \:(0.001, 1.001)$. For the WFMM, we used the same wavelet basis functions to perform DWT and implemented the MCMC procedures as described in~\citet{morris2008bayesian} to draw posterior samples. For Bayesian approaches, we ran each MCMC chain for $15000$ iterations, discarding the first $5000$ and keeping every $5$. The trace plots and Geweke diagnostic results of various parameters which are provided in the supplementary materials indicate good mixing of the chains. Using the posterior samples of $B_2^{\tau}(t)$ or $B_2^{\text{mean}}(t)$, we computed the posterior mean estimate, the $100(1-\alpha)\%$ simultaneous credible band for $\alpha \in (0,1)$ and SimBaS of the corresponding functional coefficient at each spectral location $t$. We flagged $t$ as significantly different in the $\tau$th quantile or mean between the cancer and control groups if its SimBaS is less than or equal to $0.05$ and its posterior estimate is greater than $\frac{1}{2}\log_2(1.5)$ in magnitude, corresponding to at least a $1.5$-fold change. Such flagging criteria allow us to identify regions that are both statistically and practically significant. For each non-Bayesian method, we generated $2000$ bootstrap samples to compute the mean estimate of $B_2^{\tau}(t)$ and perform functional inference. 

To perform FQR on the pancreatic dataset at each quantile level, it took about $1$ hour for Bayesian QR, $4.5$ hours for Bayesian FQR with or without adjustment, and $2.5$ hours for each bootstrap-based alternative under the computer setting specified in Section \ref{simulations}. For each quantile $\tau$, we summarized the mean estimate of $B_2^{\tau}(t)$ and the $95\%$ simultaneous credible band obtained from the Bayesian FQR and each alternative approach in plots. For the Bayesian FQR, we ran several parallel MCMC chains with different initial values at each quantile level, and obtained essentially the same point estimates and credible bands for $B_2^{\tau}(t)$. At $\tau=0.1, 0.25, 0.5$, no region was identified as significant by any of the approaches used. At $\tau=0.75, 0.9$, the regions flagged by each approach were marked on the x-axis in the corresponding plot. All these plots are available in the supplementary, and here we highlighted in Figure \ref{fig:figure4} the results for $\tau=0.9$ produced by our proposed Bayesian FQR with or without adjustment, as well as QR with wavelet denoising, an intuitive alternative that people might use to do FQR in this context, since wavelet thresholding is known to work well for spiky and spatially heterogeneous functions and in particular mass spectrometry data \citep{morris2008bayesian}. Results from FQR at $\tau=0.9$ were also compared to the functional mean regression results from WFMM in Figure~\ref{fig:figure4}.

The Bayesian FQR model with and without adjustment and the bootstrap-based QR with wavelet denoising produced an estimate of $B_2^{0.9}(t)$ that are clearly greater in magnitude than $B_2^{\text{mean}}(t)$ in the region ($5700$D, $6000$D), which coincided with what we observed from the empirical quantiles and mean in Figure~\ref{fig:figure2} (a). These quantile regression-based methods also identified far more locations than WFMM, which only flagged one narrow contiguous region [$5841.5$D, $5844.9$D]. This suggested that functional mean regression failed to detect most of the spectral locations whose protein expressions differ significantly in the $90${th} quantile between two groups.

Compared to the QR with wavelet denoising, both the Bayesian FQR and the adjusted Bayesian FQR produced much tighter $95\%$ simultaneous credible bands, allowing them to detect more locations that may correspond to proteomic biomarkers of pancreatic cancer. In particular, the Bayesian FQR flagged three contiguous regions [$5690.6$D, $5881.2$D], [$5912.4$D, $5957.7$D] and [$7607.8$D, $7619.6$D]; the adjusted Bayesian FQR flagged two contiguous regions [$5694.0$D, $5884.7$D] and [$5905.5$D, $5959.4$D]. These flagged regions covered the regions [$5704.3$D, $5789.8$D] and [$5817.4$D, $5872.6$D] flagged by the QR with wavelet denoising but included many more locations. Notably, the regions [$5912.4$D, $5957.7$D] and [$7607.8$D, $7619.6$D] were identified by our Bayesian FQR but entirely missed by the bootstrap-based approach. In addition, the bootstrap-based approach appeared to have an over-smoothed estimate of $B_2^{0.9}(t)$. For example, the Bayesian FQR detected two separate peaks at $5824$D and $5842$D, whereas the bootstrap-based approach only recognized one broader peak in this region.

The proteins corresponding to the regions flagged by our model might serve as potential biomarkers of pancreatic cancer. The expressions of these proteins differ in the $90$th quantile but not in the mean or median between the cancer cohort and the normal cohort, indicating that they are over-expressed in only a subset of cancer patients, and may fundamentally characterize unique features of this subset of pancreatic cancer patients. These potential biomarkers would have been missed by mean or median regression, with many of them missed by QR with wavelet denoising. We assessed the possible protein identities of the flagged spectral regions using TagIdent~\citep{gasteiger2005protein}, an online protein identification tool that can create a list of proteins from one or more organisms within a range of the pH and mass-to-charge ratio specified by the user. In particular, the flagged region [$5690.6$D, $5881.2$D] may correspond to basic salivary proline-rich peptide IB-$7$ ($5769$D) and peptide IB-$8$c ($5843$D) coded by PRB$2$ gene, whose single-nucleotide polymorphism (SNP) has been found to be significantly associated with the response of pancreatic cancer patients to gemcitabine based on a genome-wide association study~\citep{innocenti2012genome}. The flagged region [$5912.4$D, $5957.7$D] may correspond to a variant of transient receptor potential cation channel subfamily M member $8$ (TRPM$8$, $5940$D) which has been reported to be aberrantly expressed in pancreatic adenocarcinoma and have the potential to become a clinical biomarker and therapeutic target for pancreatic cancer~\citep{yee2012trpm7}. The narrow region [$7607.8$D, $7619.6$D] which was flagged only by our approach may correspond to stromal cell-derived factor $1$ (SDF$1$, $7610$D) coded by CXCL$12$ gene, and it has been discovered that CXCL$12$-CXCR$7$ signaling axis is significantly associated with the invasive potential of pancreatic tumor cells and the overall survival of pancreatic cancer patients~\citep{guo2016cxcl12}. To definitively find the protein identities of these spectral regions it would be necessary to conduct a tandem mass spectrometry (MS/MS) experiment~\citep{kinter2005protein,deutsch2008data}, but this is beyond the scope of our current study.

\clearpage

\begin{figure}[h]
    \centering
    \includegraphics[width=0.8\textwidth]{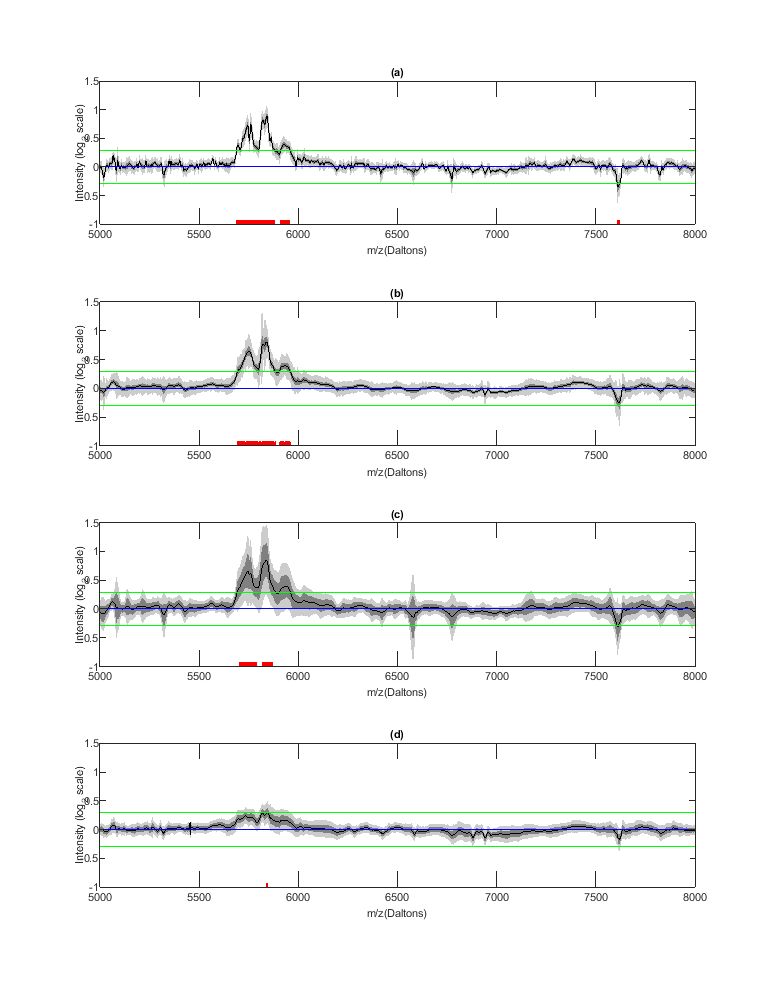}
    \vspace{-15mm}
    \caption{\textbf{\textit{Estimated cancer main effect functions for the pancreatic cancer dataset.}} (a) $B_2^{0.9}(t)$ estimated by the Bayesian FQR model. (b) $B_2^{0.9}(t)$ estimated by the adjusted Bayesian FQR model. (c) $B_2^{0.9}(t)$ estimated by the bootstrap-based QR with wavelet denoising. (d) $B_2^{\text{mean}}(t)$ estimated by the WFMM model. The estimated cancer main effects are plotted on $\log_2$ scale along with the corresponding $95\%$ simultaneous credible bands. A spectral location is flagged as significant and marked on the x-axis if its SimBaS is less than or equal to $0.05$ and the estimate corresponds to at least $1.5$-fold change indicated by the two horizontal lines.}
    \label{fig:figure4}
    \vspace{-5mm}
\end{figure} 

\clearpage

\section{Discussion} \label{discussion}
In this paper, we introduced a fully Bayesian approach to perform quantile regression on functional responses. The existing work on functional response regression has focused predominantly on mean regression. However, sometimes predictors may not strongly influence the conditional mean of functional responses, but other aspects of their conditional distributions instead, as illustrated by our analysis of the motivating pancreatic cancer mass spectrometry dataset. In this case, performing functional quantile regression to delineate the relationship between functional responses and predictors is warranted. This can straightforwardly be done by performing quantile regression at each individual functional location, but as we demonstrate this is not an efficient strategy since it fails to borrow strength from nearby functional locations. Our proposed approach borrows strength across nearby locations by representing the functional coefficients with appropriate basis functions, and induces adaptive penalization on the basis coefficients by placing a global-local shrinkage prior. We developed a scalable data augmented block Gibbs sampler for posterior computation, which can be implemented automatically without tuning parameters and scale up well to moderately-sized functional data consisting of hundreds of observations per curve. Posterior samples were used to perform Bayesian estimation and inference on parameters of interest while accounting for multiple testing. In the pancreatic cancer data application, our Bayesian FQR model identified many more spectral locations compared to mean-based alternatives, which correspond to proteins whose intensity levels differ significantly in the $90${th} quantile but not the mean between the cancer and normal populations. 

Our framework is flexible in that it allows different types of basis transform and continuous shrinkage priors, which are chosen based on the characteristics of functional data. We chose to use wavelets and a horseshoe prior to present our approach, which are well-suited for the highly spiky and irregular mass spectrometry data. Other basis functions including functional principal components, Fourier series and splines and a great variety of shrinkage priors can also be used, as elaborated in Section \ref{model}. In addition, our framework can accommodate multi-dimensional functional data by applying a multi-dimensional basis transform. For example, a $2$D wavelet transform can be applied to the $2$D mass spectrometry data collected in LC-MS experiment~\citep{zhang2009review,liao2014new}. % so this approach can be used to perform FQR on data from these assays as well.
We assumed the conditional quantile to be linear in the covariates in this paper, but our model can be easily extended to model nonparametric effect of covariates \citep{kim2007quantile, cai2008nonparametric, wang2009quantile, feng2016estimation, fasiolo2018fast} by using spline design matrices.

We simulated functional data with Gaussian shaped peaks to mimic mass spectra, evaluated the performance of our method and compared to simpler alternatives that people might use to perform FQR in the simulation study. Our approach consistently outperformed the na\"ive Bayesian quantile regression in both estimation and inference, showing that it is inefficient to ignore the functional nature of data and do quantile regression separately for each location. In addition to borrowing strength, our model adopted a sparsity prior that can effectively shrink small wavelet coefficients to zero and avoid attenuation of large coefficients, minimizing bias and substantially reducing variation in parameter estimation.

We also considered bootstrap-based two-step alternatives, which are themselves new methods that we introduced to compare with our proposed approach. One might think of it a natural approach to draw bootstrap samples of observed functional data and post-smooth the pointwise quantile regression estimates in each bootstrap iteration, using spline smoothing or wavelet denoising. Compared to these two-step alternatives that seemed intuitively appealing, our approach achieved comparable estimation accuracy but considerably smaller variability, which led to much tighter simultaneous credible band with similar coverage, and greatly improved sensitivity for identifying significant regions in the functional coefficients at particular quantile levels. This improvement of our Bayesian FQR model could be explained by the fact that quantile regression and penalization of functional coefficients are performed jointly in a unified manner in our Bayesian framework. The possibly heteroscedastic noise levels across $t$ in the functional data are learned in the quantile regression step and then carried forward to the coefficient penalization step, which we believe to have the potential to achieve more adaptive regularization than performing them separately as done in the two-step approaches. While our Bayesian hierarchical model is convenient to implement, it would be very challenging to fit a non-Bayesian counterpart with the same flexibility and complexity, and yield estimation and inference of $\mathbf{B}$ while choosing various penalization parameters $\lambda_{ajh}$ and $\psi_{aj}$ by cross-validation.

We chose to use the asymmetric Laplace likelihood as the working likelihood in our framework due to its computational efficiency. Motivated by recent studies raising concerns about the frequentist propeties of posterior inference based on this likelihood, we also considered an adjusted Bayesian FQR model by performing a pointwise likelihood correction proposed by \citep{sriram2015sandwich}, and compared its performance to our original model in simulation studies and data application. The simulation results showed that the original model had satisfactory performance in terms of parameter estimation and signal detection in all the scenarios we considered; the adjustment procedure further improved estimation accuracy and led to slightly wider credible bands, and essentially removed any false positives at the expense of slightly decreased sensitivity compared to the original model. While our adjustment is done separately at each individual location and seems ad-hoc, it does have very good empirical performance based on our simulations. It would be insightful to extend this adjustment strategy to the functional data setting so that it can accommodate the within-function dependence structure and also to study its asymptotic properties, but these are beyond the scope of our current work.

There exists limited work on FQR in the literature. Based on our simulations, the framework proposed by \citet{brockhaus2015functional} appears to work satisfactorily for simple and homogeneous functions sampled on a relatively sparse grid, but not as well for high-dimensional spiky and complex functions in terms of coefficient estimation and computational feasibility. In addition, their framework does not automatically yield pointwise or joint inference.

One should always ensure that the effective sample size $N \min\{\tau, 1-\tau\}$ is sufficiently large before performing FQR at $\tau$th quantile. While we propose a highly flexible and computationally tractable Bayesian framework to perform FQR, there is still room for improvement. Our modeling approach is built for functional data sampled on a sufficiently fine grid where interpolation can be reasonably used to obtain a common grid for subjects. Further adaptations of our model would be required for functional data sampled on sparse grids that vary across subjects. We assume independent residual errors across $t$, but observations from nearby functional locations are typically correlated. This independent error assumption may lead to conservative inference, thus further efficiency and power gains are possible if within-function correlations could be accommodated~\citep{morris2017comparison}. However, the tractability of our proposed framework breaks down if we are to model this dependence structure. While it is relatively easy to account for intrafunctional correlations in functional mean regression, we find it much more challenging to do so for FQR, which has never yet been addressed in the existing literature to our best knowledge. It should be pointed out that even with an independent error assumption, our proposed approach still beats all the simpler methods that people might use to perform FQR as shown by the simulations, so we believe our work is a significant step forward in this area. Finally, alternative regularization methods on the basis coefficients can be explored, such as the FLiRTI model~\citep{james2009functional} that enforces sparsity in the functional coefficients or their derivatives to improve interpretability.

\bigskip

\begin{center}
{\large SUPPLEMENTARY MATERIAL}

The supplementary materials include mathematical details of the MCMC sampling procedure and additional results of mass spectrometry data application. The pancreatic cancer mass spectrometry dataset, simulation datasets and the related MATLAB and R code are available at \href{https://github.com/MorrisStatLab/FunctionalQuantileRegression}{https://github.com/MorrisStatLab/FunctionalQuantileRegression}.
\end{center}

\bigskip

\bibliographystyle{Chicago}
\bibliography{references}

\end{document}